\documentclass[reprint,eqsecnum,floats,aps,amsmath,amssymb,nofootinbib,prd,onecolumn, showpacs]{revtex4-2}
\usepackage{graphicx,physics}
\usepackage{amsmath,amssymb,mathtools,mathrsfs}
\usepackage{hyperref}
\usepackage{graphicx}
\usepackage{subfigure}
\usepackage{arydshln}
\usepackage{xcolor}
\usepackage{braket}
\usepackage{tensor}
\usepackage{enumitem,array,textcomp}
\usepackage[utf8x]{inputenc}

\begin{document}
	
	\title{Mass of cosmological perturbations in the hybrid and dressed metric formalisms of Loop Quantum Cosmology for the Starobinsky and exponential potentials}
	\author{Simon Iteanu}
 	\email{simon.iteanu@ens-lyon.fr}
	\affiliation{Ecole Normale Sup\'erieure de Lyon, 15 parvis René Descartes BP 7000, 69342 Lyon Cedex 07, France; Instituto de Estructura de la Materia, IEM-CSIC, Serrano 121, 28006 Madrid, Spain}
	\author{Guillermo  A. Mena Marug\'an}
	\email{mena@iem.cfmac.csic.es}
	\affiliation{Instituto de Estructura de la Materia, IEM-CSIC, Serrano 121, 28006 Madrid, Spain}
	
\begin{abstract}
The hybrid and the dressed metric formalisms for the study of primordial perturbations in Loop Quantum Cosmology lead to dynamical equations for the modes of these perturbations that are of a generalized harmonic-oscillator type, with a mass that depends on the background but is the same for all modes. For quantum background states that are peaked on trajectories of the effective description of Loop Quantum Cosmology, the main difference between the two considered formalisms is found in the expression of this mass. The value of the mass at the bounce is especially important, since it is only in a short interval around this event that the quantum geometry effects on the perturbations are relevant. In a previous article, the properties of this mass were discussed for an inflaton potential of quadratic form, or with similar characteristics. In the present work, we extend this study to other interesting potentials in cosmology, namely the Starobinsky and the exponential potentials. We prove that there exists a finite interval of values of the potential (which includes the zero but typically goes beyond the sector of kinetically dominated inflaton energy density) for which the hybrid mass is positive at the bounce whereas the dressed metric mass is negative.
\end{abstract}

	
\maketitle					

\section{Introduction}\label{sec:intro}

Loop Quantum Gravity (LQG) \cite{ALQG,Thiem} is a nonperturbative and background independent approach for the quantization of general relativity, proposed by Ashtekar approximately thirty-five years ago \cite{asht,asht2}. In LQG, the geometry is described in terms of fluxes of densitized triads and holonomies of their conjugate $su(2)$-connections. To put the theory to the test and search for physical consequences that might eventually be detected, LQG has been applied to symmetry reduced spacetimes, in a discipline that is nowadays called Loop Quantum Cosmology (LQC) \cite{AS,LQCG,aps,aps2,MMO}. In particular, in the last ten years there has been considerable activity in the field of cosmological perturbations, trying to extract predictions about imprints of quantum geometry effects in diverse cosmological observables, with a prominent role played by the power spectra of the cosmological microwave background (CMB) (a list of works that, although not exhaustive, covers different approaches to this issue, with an emphasis on those on which we are going to focus our attention, is given by Refs. \cite{hybrid1,hybrid2,hybrid3,hybrid4,hybrid5,hybrid6,hybrid7,dressed1,dressed2,dressed3,dressed4,
Wang1,Wang2,Wang3,Wang4,tangoPRL,tango,dressedIvan,effective1,effective2,effective3,Edward,hybridreview}). 
More recently (and after some pioneering works, see e.g. Refs. \cite{bh1,bh2}), there have been different attempts to extend the scope of LQC to black hole spacetimes (some representative publications are Refs. \cite{lqcbh1,lqcbh2,lqcbh3,lqcbh4,lqcbh5,lqcbh6,lqcbh7,lqcbh8,AOS,AOS2,AO,AG,AG2,ABG}). In this front, an eye is put on the possible traces of LQG on the gravitational radiation emitted by black holes. 

Among the various proposals developed to handle the influence of LQC on primordial perturbations in the very early universe, there are two that respect the hyperbolic ultraviolet behavior of the propagation equations for the perturbations. They are the so-called dressed metric formalism \cite{dressed1,dressed2,dressed3} and the hybrid LQC formalism \cite{hybridreview}. Moreover, they both lead to ignorable modifications on the dynamics of the perturbations for wavelengths much smaller than the typical curvature scale of the era of large quantum geometry effects \cite{dressed4,universe}, so that they reproduce fairly well the standard cosmological predictions in that sector of wavelengths. If the curvature scale lays today in the observable region of large angular scales, then the two proposals are compatible with the observations of the CMB while introducing corrections precisely only in that part where recent records suggest the presence of tensions with the standard theory, or at least of statistical anomalies \cite{tango,dressedIvan,hybridreview}.

In the two discussed formalisms one usually considers perturbations with a negligible backreaction on the cosmological background. This is required for consistency in the case of the dressed metric \cite{dressed3}, and it is often the situation that is studied in the hybrid formalism (where one could cope in principle with some backreaction \cite{hybridreview}). In these scenarios with ignorable backreaction, the dynamical equations that rule the propagation of the primordial perturbations are very similar in the two formalisms. For each (Fourier) mode of the perturbations, the dynamics can be described as a generalized harmonic-oscillator equation with a background-dependent mass\footnote{For the sake of brevity, we will refer to the mass term in the dynamical equations of the perturbations as the mass, although strictly speaking it corresponds to a square mass. Given the well-delimited context of our discussion, this terminology should not lead to any misunderstanding.} that is the same for all modes \cite{mass}. The only difference between the two approaches is the exact expression for this mass \cite{mass,hybridreview}, owing to the different quantization prescription followed in each case. 

Furthermore, it is common to consider quantum states for the background geometry that stay peaked on the evolution about a trajectory of an effective Hamiltonian, $H_0^{\rm eff}$, which differs from the classical one for the cosmological background in general relativity, $H_0$, by the inclusion of quantum modifications \cite{aps2,taveras}. On these trajectories, the big bang singularity is replaced with a bounce (namely, the so-called big bounce). The background dependence of the mass of the perturbations can then be estimated, in a good approximation, by evaluating it on the peak of the quantum state. Even in this effective regime, the expression of the mass differs for the dressed metric and the hybrid formalisms \cite{mass}. The reason behind this discrepancy is the following. In the dressed metric formalism, the time derivatives that appear are computed using the effective Hamiltonian (which is the generator of the evolution on the peak trajectories). On the other hand, in the hybrid formalism, time derivatives are expressed in terms of canonical variables (configuration and momenta) before quantization and their restriction to the effective regime. It is only the final expression in terms of those canonical variables that is evaluated effectively. The result is not the same as for the dressed metric, because taking Poisson brackets (or commutators) and evaluating on effective trajectories are not processes that commute \cite{mass}. To see this in more detail, let us consider, in a homogeneous and isotropic cosmology, the second derivative of geometric quantities, such as the scale factor $a$, with respect to conformal time. This derivative is given classically by the Poisson brackets $a\{a\{\cdot,H_0\},H_0\}$. For quantum states peaked on effective trajectories of LQC, these brackets become $[a\{a\{\cdot,H_0\},H_0\}]_{\rm eff}$ and $[a\{a\{\cdot,H_0^{\rm eff}\},H_0^{\rm eff}\}]_{\rm eff}$, respectively in the hybrid and the dressed metric formalisms \footnote{It is clear why the Hamiltonian $H_0$ appears in these brackets in the hybrid formalism while the corresponding Hamiltonian is $H_0^{\rm eff}$ in the dressed metric case. The hybrid formalism rests on a purely canonical formulation, for which the background dynamics is dictated by $H_0$. On the other hand, the dressed metric approach first {\sl dresses} the background metric with quantum corrections and then lifts the resulting dynamics to the truncated phase space that describes the perturbations (see the original formulation of the dressed metric formalism in Ref. \cite{dressed2}). The dynamical evolution of this dressed metric is ruled by the effective Hamiltonian, $H_0^{\rm eff}$. See also Ref. \cite{Ref0} for an alternative viewpoint on the differences and relation between these two formalisms.}. Here, the subscript ${\rm eff}$ indicates evaluation on effective solutions after having computed the brackets. Remarkably, these two double Poisson brackets do not coincide. Then, recalling that the classical mass term is supplied by a second derivative in conformal time (e.g., of $a$ in the case of tensor perturbations), we conclude that the hybrid and the dressed metric masses necessarily differ when the effective dynamics departs from the genuinely classical one.

Since relevant quantum geometry effects occur only in a narrow period around the big bounce \cite{hybridreview,analyticvacuum}, it is precisely in this region where the difference between the masses obtained with the two quantization approaches is not ignorable. In particular, the positivity of one or the other of these masses at the bounce is of the greatest importance in order to define mode solutions of positive frequency and implement proposals for the specification of the vacuum of the perturbations, precisely in the very period where we believe that quantum effects are significant (details about these questions will be presented in in Sec. \ref{sec:conclusion}). Because of this, the analysis of the background-dependent mass of the perturbations at the bounce is especially interesting. In a previous paper, the difference between the mass of the dressed metric and hybrid formalisms was studied at the bounce for tensor and scalar perturbations \cite{mass}. It was proved that, while the mass at the bounce of hybrid LQC is positive for a finite interval of values of the potential that includes the sector corresponding to kinetically dominated inflaton energy density, the mass of the dressed metric approach is negative in an interval with the same characteristics.  The discussion in that work was particularized to inflaton potentials with the properties of a mass contribution, namely, positive potentials with a second derivative that can be treated as a parameter (the inflaton mass), and satisfying a bound on the ratio of its first derivative squared and the product of the second derivative with the potential itself (this bound becomes in fact an identity in the case of a mass contribution). In particular, these properties were employed in the analysis of the mass of the scalar perturbations. 

The aim of this work is to extend this discussion to other potentials that are interesting in cosmology. We will study the Starobinsky potential \cite{staro} and a family of exponential potentials that contains the case of the purely exponential and the hyperbolic cosine. 

The Starobinsky potential has been suggested in the framework of the standard cosmological model as an alternative to the simplest possibility of a quadratic potential corresponding to a mass contribution. A statistical analysis of the observational data (see e.g. the results of the Planck mission in Refs. \cite{planck15,planck18}) has shown that the Starobinsky potential provides a much better fitting. This indicates a preference for the Starobinsky potential in the case of the standard inflationary model within general relativity. In principle, this preference cannot be directly extrapolated to the study of perturbations in LQC, case in which a similar statistical analysis of the data has not been conducted, including different families of potentials and adapting the choice of vacuum state to the new situation \cite{universe,analyticvacuum}, instead of adopting the standard Bunch-Davies vacuum \cite{bunch} as for slow-roll inflation. Nevertheless, a considerable number of works have studied the Starobinsky potential in LQC \cite{Wang1,Wang2,Wang3,Wang4,tangoPRL,tango,Bonga1,Bonga2,louk}, with the hope that the alleviation of the statistical tensions will be even more favorable when combining this potential with the presence of small quantum corrections. Therefore, discussing the behavior of the background-dependent mass of the perturbations for the Starobinsky potential is especially relevant, including the determination of its positivity or negativity at the bounce.

The exponential potentials, on the other hand, have a more academic interest. A purely exponential potential is known to lead to power-law inflation \cite{hall,luchin,ratra}. At least in geometrodynamics, a generic linear combination of the exponential of the inflaton and the exponential of its negative can always be reformulated as a problem with a purely exponential potential or with a hyperbolic cosine potential after suitable field redefinitions \cite{halliwell}. Moreover, again in the context of geometrodynamics, the background cosmology obtained with these two specific potentials turns out to be exactly solvable, not only classically, but also quantum mechanically \cite{halliwell}. It would be interesting to explore if this solvability can be maintained somehow in the loop quantization, or at least in the effective loop dynamics. As an important step towards the consideration of perturbations in this framework, we will discuss here the properties of the background-dependent mass, with the focus on the big bounce, around which all non-ignorable quantum effects occur.

The rest of this work is organized as follows. We first succintly review in Sec. \ref{sec:mass} the formulas for the background-dependent mass of the tensor and scalar perturbations in the effective regime for the hybrid and dressed metric formalisms of LQC, particularizing them at the bounce. We also introduce there our conventions, and the expressions of the Starobinsky and exponential potentials. We then study the behavior of the mass for the Starobinsky potential in Sec. \ref{sec:starobinsky}, proving its positivity for the hybrid approach and its negativity for the dressed metric approach in a finite interval of values of the potential that includes the origin. This interval contains the sector corresponding to kinetic dominance in the inflaton energy density, for very small potentials, but the conclusion is much more robust in the sense that it typically applies as well beyond this sector, for values of the potential that can exert an influence on the evolution. A similar study with comparable conclusions is obtained for the exponential potentials in Sec. \ref{sec:expo}. We study in detail the hyperbolic cosine potential and the purely exponential potential. We finally present our conclusions and some further comments in Sec. \ref{sec:conclusion}. We adopt natural units in which the speed of light and the reduced Planck constant are equal to one. Planck units are then defined by taking Newton constant also equal to one.

\section{The mass} \label{sec:mass}

We consider a cosmological background consisting of a spacetime with homogenous and isotropic flat spatial sections, with metric described by the scale factor $a$ (or by a triad variable proportional to its square \cite{aps}), and a homogenous scalar field $\phi$ subject to a potential $V(\phi)$. This field is able to produce an inflationary expansion during the cosmological evolution, and in this sense we refer to it as the inflaton. The energy density $\rho$ and the pressure $P$ of this inflaton are
\begin{equation}\label{density}
\rho=\frac{1}{2}\left(\frac{\phi'}{a}\right)^2+V(\phi) ,\qquad P=\rho-2 V(\phi),
\end{equation}  
where the prime denotes the derivative with respect to conformal time. These expressions are valid not only for classical backgrounds, but also for quantum backgrounds that are peaked on trajectories of the effective LQC Hamiltonian. For states of this kind that experience a bounce, the inflaton energy density takes its maximum at that moment, equal to $\rho_{\rm max}=3/(8\pi G \gamma^{2}\Delta)$, which is approximately of Planck order, roughly speaking. Here, $\gamma$ is the so-called Immirzi parameter \cite{immi} and $\Delta=4\sqrt{3}\pi\gamma G$ is the area gap, namely the minimum nonzero area allowed by the spectrum of the area operator in LQG \cite{Thiem}. 

In this work, we will discuss in detail two types of inflaton potentials. The first one is the Starobinsky potential, that has the form
\begin{equation}\label{starobinsky}
V(\phi)=V_S(\phi):=\frac{3 M^2}{32\pi G^2} \left(1- e^{- D_S \phi} \right)^2,
\end{equation}
where $M$ and $D_S$ are constants. The latter can be chosen to be positive, absorbing its sign with a flip of sign in $\phi$ if necessary. One usually sets $D_S=\sqrt{16\pi G/3}$ (with a suitable scaling of the inflaton). On the other hand, and although we will treat $M$ as a free parameter, we will assume that it does not deviate much from its most favored value according to the observations of the Planck mission \cite{planck15}, which is $M = 2.51 \times 10^{-6}$. To simplify our notation, we will call $A_S^2=3 M^2/(32\pi G^2)$, where we have made use of the positivity of this constant. Let us mention that the potential $V_S(\phi)$ can be obtained from a higher-order gravitational theory with quadratic curvature corrections by means of a conformal transformation of the metric and an adequate definition of the scalar field $\phi$ \cite{conf1,conf2,conf3,conf4}.

The other type of potential that we will study is given by linear combinations of the exponentials of plus or minus $D\phi$, where $D$ is a constant. We will focus on two important cases (that at least in geometrodynamics cover all possible cases after a redefinition of variables), namely, a hyperbolic cosine and a pure exponential, with the potential respectively given by
\begin{equation}\label{expopotentials}
V(\phi)= V_C(\phi):=A_C \cosh{(D_C\phi)},\qquad V(\phi)=V_e(\phi):=A_e e^{D_e\phi}.
\end{equation} 
Here, $A_C$, $D_C$, $A_e$, and $D_e$ are constants. We will treat their values (in Planck units) as parameters. We will restrict our attention to positive potentials, which is the case of greatest physical interest. This, together with the invariance under parity of the hyperbolic cosine and the possibility of absorbing the sign of $D_e$ with a flip of sign in the inflaton, leads us to consider only positive constant parameters in the above potentials. On the other hand, we will allow the possible addition of a constant $E_C$ to the coshine potential, namely $V(\phi)=\tilde{V}_C(\phi):=V_C(\phi)+E_C$, so that its minimum, which is reached at vanishing inflaton, can be set equal to zero by choosing $E_C=-A_C$. 

The perturbations of both the inflaton and the geometry propagate on this cosmological background. The physical part of those perturbations is captured in gauge invariants, which do not change under perturbative spacetime diffeomorphisms \cite{hybridreview}. Given our cosmological system, the only gauge invariant perturbations can be identified as one scalar and two tensor perturbations (the latter corresponding to two different polarizations) \cite{hybridreview}. In our flat background, the scalar perturbation can be conveniently described in terms of the Mukhanov-Sasaki gauge invariant \cite{sasa,kodasasa,mukhanov}, which can be directly related with the comoving curvature perturbations that induce the temperature anisotropies. 

Expanding all gauge invariant perturbations in Fourier modes, and after a suitable background-dependent redefinition of their Fourier coefficients, one can show that their equations of motion in the effective regimes of both hybrid and dressed metric LQC can be written as generalized harmonic-oscillator equations with a background-dependent mass, as we have already commented \cite{mass}. For instance, for a mode  of the Mukhanov-Sasaki field with wavevector $\vec{k}$, described by the time-dependent mode coefficient $v_{\vec{k}}$, the resulting equation in the hybrid formalism of LQC is 
\begin{equation}\label{MSequ}
v_{\vec{k}}^{\prime\prime}+[k^2+s^{({\rm s})}]v_{\vec{k}}=0,
\end{equation}
where $k$ is the wavenorm and $s^{({\rm s})}$ is the (square) mass for the scalar perturbations, which we will call the scalar mass. In addition, let $s^{({\rm t})}$ be the corresponding tensor mass in this hybrid formalism. Similarly, let $\breve{s}^{({\rm t})}$ and $\breve{s}^{({\rm s})}$ denote the respective counterparts of the tensor and scalar masses in the dressed metric approach. We recall that none of these masses depends on the Fourier mode under study (nor on its polarization, in the tensor case) \cite{hybridreview}. 

These masses adopt the expressions \cite{mass}
\begin{eqnarray}\label{hmass}
s^{({\rm t})}&=&-\frac{4\pi G}{3}a^2(\rho-3P),\qquad s^{({\rm s})}=s^{({\rm t})}+\mathcal{U},\\ \label{dmass}
\breve{s}^{({\rm t})}&=&-\frac{4\pi G}{3}a^2\rho\left(1+2\frac{\rho}{\rho_{\rm max}}\right)+4\pi G a^2 P\left(1-2\frac{\rho}{\rho_{\rm max}}\right), \qquad \breve{s}^{({\rm s})}=\breve{s}^{({\rm t})}+\mathcal{V}.
\end{eqnarray}
The contributions $\mathcal{U}$ and $\mathcal{V}$ provide the difference between the tensor and scalar masses in the hybrid and the dressed metric formalisms, respectively, and become zero for a vanishing potential. They are given by 
\begin{eqnarray}\label{Ueff}
\mathcal{U}&=&a^2\left[V_{,\phi\phi}+48\pi G V(\phi)-\frac{48\pi G}{\rho} V^2(\phi)+ 6\frac{a^{\prime}{\phi}^{\prime}}{ a^{3}\rho}V_{,\phi}\right],\\ \label{Veff}
{\mathcal{V}}&=&a^2\left[V_{,\phi\phi}+48\pi G V(\phi)-\frac{48\pi G}{\rho} V^2(\phi)-  \sigma_a \sqrt{\frac{96\pi G}{\rho}} \frac{|\phi^{\prime}|V_{,\phi}}{a}\right].
\end{eqnarray}
A subscript in $V$ with a comma followed by the symbol $\phi$ denotes the derivative of the potential with respect to the inflaton (once or twice, depending on the number of times that $\phi$ appears). Besides, the symbol $\sigma_a$ stands for the sign of the product of $\phi^{\prime}$ with the canonical momentum of the scale factor, evaluated on the considered effective trajectory \cite{mass}. In our calculations, we will just need that it can only take the values $\pm1$.

At the bounce, the inflaton energy density becomes equal to $\rho_{\rm max}$ and the scale factor reaches its minimum, which we call $a_{\rm B}$. We will also use the symbol ${\rm B}$ as a subscript or superscript in any of our variables to indicate evaluation at the bounce. For convenience, we will also use it as a superscript in the potential and its derivatives to denote evaluation at the value of the inflaton at the bounce, $\phi_{\rm B}$. At such instant of time, our formulas for the mass simplify significantly. We get \cite{mass}:
\begin{eqnarray} \label{hybriddressedformula}
\frac{s^{({\rm t})}_{\rm B}}{8\pi G a_{\rm B}^2}&=&\frac{\rho_{\rm max}}{3}- V^{\rm B}, \qquad 
\frac{\breve{s}^{({\rm t})}_{\rm B}}{8\pi G a_{\rm B}^2}=-\rho_{\rm max}+ V^{\rm B}, \\ \label{Uformula}
\frac{{\mathcal{U}}_{\rm B}}{8\pi G a_{\rm B}^2}&=&  \frac{1}{8\pi G}\left[V_{,\phi\phi}^{\rm B}+48\pi G V^{\rm B}-\frac{48 \pi G}{ \rho_{\rm max}} \left(V^{\rm B}\right)^2\right],\\ \label{Vformula}
\frac{{\mathcal{V}}_{\rm B}}{8\pi G a_{\rm B}^2}&=& \frac{{\mathcal{U}}_{\rm B}}{8\pi G a_{\rm B}^2} - \sigma_a \sqrt{\frac{3} {2 \pi G \rho_{\rm max}} } \frac{|{\phi}'_{\rm B}|V_{,\phi}^{\rm B}}{a_{\rm B}}.
\end{eqnarray}

It is straightforward to see that the tensor mass at the bounce is positive in hybrid LQC for values of the potential smaller than $\rho_{\rm max}/3$ (including the zero), whereas it is always negative in the dressed metric formalism. This last statement follows from the fact that $\rho_{\rm max} - V^{\rm B} $ is just the kinetic contribution to the energy density at the bounce, which is always equal or greater than zero. This is consistent with the well-known fact that the tensor mass coincides with the effective value of $-a''/a$ in the dressed metric formalism, and the second derivative of the scale factor is positive at the bounce, because $a_{\rm B}$ is a minimum \cite{mass}. Notice that the proven positivity or negativity of the tensor mass applies well beyond the sector of kinetically dominated energy density, in which the potential is negligible. Our conclusions are much more robust than a result exclusively for the kinetic regime, since they extend to regions where the potential takes values that cannot be ignored in the analysis. In the two following sections we will study the behavior of the scalar mass at the bounce. For simplicity in our formulas, and to allow a treatment of the parameters of the potential as mere numbers equal to their Planck values, from now on we adopt Planck units, setting $G=1$.

\section{The Starobinsky potential} \label{sec:starobinsky}

Let us start by considering the scalar mass at the bounce in the case of the Starobinsky potential, with value $V_S^{\rm B}=A_S^2 (1- e^{-D_S \phi_{\rm  B}})^2$. It is convenient to introduce the notation $y = A_S e^{-D_S \phi_{\rm B}}$, with $A_S>0$, so that $y$ is greater than zero. When the inflaton at the bounce runs over the real line, $y$ can take any positive real value, regardless of $A_S$. In terms of this variable, the potential and its derivatives can be expressed as
\begin{equation}\label{starobounce}
V^{\rm B}_S = (y - A_S )^2, \qquad V_{S,\phi}^{\rm B} = -2D_S  (y - A_S) y, \qquad V_{S,\phi\phi}^{\rm B} = 2 D_S^2  (2y - A_S) y.
\end{equation}
Since, in the phenomenologically interesting case for cosmological perturbations, $A_S$ is of the order of $10^{-6}$--$10^{-7}$ in Planck units \cite{planck15}, in the following we restrict our analysis to values of $A_S$ such that this constant can be treated as a perturbative parameter in the potential. We divide our discussion in two parts. First, we study the mass at the bounce for the hybrid formalism, and then we consider the mass for the dressed metric formalism of LQC.

\subsection{Starobinsky potential in the hybrid formalism}\label{subsec:hybstarobinsky}

The mass for the Starobinsky potential in the hybrid formalism is given at the bounce by 
\begin{eqnarray}
\frac{s_{\rm B}^{({\rm s})}}{8\pi a^2_{\rm B}}&=& \frac{\rho_{\rm max}}{3} + 5 A_S^2 - \frac{6 A_S^4 }{\rho_{\rm max}} - A_S \left(10  + \frac{D^{2}_S}{4 \pi}- \frac{24 A_S^2} {\rho_{\rm max}} \right)  y  \nonumber \\ \label{hybridmassstaro}
&+& \left( 5 + \frac{D^2_S}{2\pi}- \frac{36 A_S^2}{ \rho_{\rm max}}\right) y^2 
+ \frac{24 A_S}{ \rho_{\rm max}} y^3 - \frac{6} { \rho_{\rm max}} y^4.
\end{eqnarray}
At dominant, zeroth order in $A_S$, this quartic polynomial in $y$ can be approximated by
\begin{equation} \label{polhibridmassstaro}
P(y) = \frac{\rho_{\rm max}}{3} + \left(5  + \frac{D_S^2}{2\pi}\right) y^2 - \frac{6}
{\rho_{\rm max}} y^4.
\end{equation}
This quadratic polynomial in $y^2$ has a positive discriminant that is greater than the square of the coefficient of $y^2$. Hence, the polynomial admits a unique positive root, and therefore the same occurs in terms of the original variable $y>0$. This root is  
\begin{equation}\label{roothybstaro}
y_0 =\left( \frac{10\pi+D_S^2 + \sqrt{ 132 \pi^2 + 20 \pi D_S^2  + D_S^4 }  } {24\pi} \rho_{\rm max} \right)^{1/2}.
\end{equation}
It is straightforward to see that 
\begin{equation}
y^2_0 \geq \frac{5+ \sqrt{33}}{12} \rho_{\rm max} \approx 5.372 \times \frac{\rho_{\rm max}}{6}.
\end{equation} 
Therefore, we conclude {\sl a fortiori} that $s_{\rm B}^{({\rm s})}$ is positive (at dominant order in $A_S$) for all $V_S^{\rm B} \leq 5 \rho_{\rm max}/6$. Clearly, this interval of values covers the case of a kinematically dominated scenario, in which the potential should not exceed a small fraction of $\rho_{\rm max}$. 

Finally, note that we can improve our approximation at zeroth order in $A_S$ by considering higher orders in this parameter. At next order, for instance, we can search for a root of the form $\tilde{y}_0=y_0+A_S \,y_1$ to the linear truncation in $A_S$ of the right-hand side of Eq. \eqref{hybridmassstaro}. After a straightforward calculation, we obtain
\begin{equation}\label{y1hybstaro}
y_1 = 1 + \frac{3D_S^2}{4\sqrt{132 \pi^2 + 20 \pi D_S^2  + D_S^4 }} . 
\end{equation}
In a similar way, we can recursively increase the order of our approximation beyond the linear truncation in $A_S$. 

\subsection{Starobinsky potential in the dressed metric formalism}\label{subsec:dressstarobinsky}

In the dressed metric formalism, the analysis is complicated by the appearance of an additional term in the mass at the bounce that is proportional to the derivative of the potential, $V_{S,\phi}^{\rm B}$. The expression of this mass is
\begin{eqnarray}
\frac{\breve{s}^{({\rm s})}_{\rm B}}{8\pi a^2_{\rm B}}&=& - \rho_{\rm max} + 7 A_S^2 - \frac{6 A_S^4}{\rho_{\rm max}} -  A_S  \left(  14 + \frac{D^2_S}{ 4\pi} - \frac{24 A_s^2}{ \rho_{\rm max}} \right) y + \left( 7 + \frac{D^2_S}{2 \pi} - \frac{36 A_S^2}{\rho_{\rm max}}\right) y^2  \nonumber \\ \label{dressedmassstaro}
&+&  \frac{24 A_S}{ \rho_{\rm max}} y^3 - \frac{6}{ \rho_{\rm max}} y^4 \pm  \sqrt{\frac{3}{2\pi \rho_{\rm max}} }  \frac{|\phi^{\prime}_{\rm B}|V_{S,\phi}^{\rm B}}{a_{\rm B}}  , 
\end{eqnarray}
where we have taken into account that the sign $\sigma_a$ in Eq. \eqref{Vformula} can be equal to $\pm 1$. 

We can bound the mass at the bounce from below and above with two polynomials of $V_S^{\rm B}$ by finding a suitable bound for the factor $|\phi^{\prime}_{\rm B}| V_{S,\phi}^{\rm B}/a_{\rm B}$. First, from the expression of the inflaton energy density [see Eq. \eqref{density}], we notice that 
\begin{equation}
|\phi^{\prime}_{\rm B}|=a_{\rm B}  \sqrt{ 2(\rho_{\rm max} - V_S^{\rm B})}\leq a_{\rm B} \sqrt{2 \rho_{\rm max} } ,
\end{equation} 
the last inequality following from the positivity of the Starobinsky potential. Then, using the form of the derivative of this potential at the bounce, we get $V_{S,\phi}^{\rm B} \approx -2 D_S y^2$ at zeroth order in the negligible parameter $A_S$, with the same notation for $y$ as before. In this way and recalling that $D_S$ is positive, we conclude that, at dominant order,
\begin{equation}\label{boundstaro}
\frac{|\phi^{\prime}_{\rm B} V_{S,\phi}^{\rm B}|}{a_{\rm B}}  \leq 2 D_S \sqrt{2 \rho_{\rm max}} y^2 .
\end{equation}

Using the above inequality, it is straightforward to see that, at a zeroth-order truncation in the parameter $A_S$, the following polynomials bound from above $(+)$ and below $(-)$ the mass at the bounce for the Starobinsky potential in the dressed metric formalism:
\begin{equation}\label{pMmstaro}
P_{\pm}(y)= - \rho_{\rm max} + \left( 7  \pm 2  \sqrt{ \frac{3}{ \pi} }D_S + \frac{D^2_S}{ 2\pi}\right)  y^2 - \frac{6}{ \rho_{\rm max}} y^4. 
\end{equation}
Concretely, 
\begin{equation}
P_{-}(y) \leq \frac{\breve{s}^{({\rm s})}_{\rm B}}{8\pi a^2_{\rm B}}\leq P_{+}(y).
\end{equation}

Let us study $P_{+}(y)$ in particular. Regarded as a quadratic function of $y^2$, its discriminant $\Delta P_{+}$ is
\begin{equation}\label{DeltapMstaro}
\Delta_{P_{+}}= \left( 7 + 2  \sqrt{ \frac{3}{ \pi} }D_S+ \frac{D^2_S}{ 2\pi} \right)^2 - 24.
\end{equation}
We see that it is always strictly positive and smaller than the square of the term in parentheses in the right-hand side. As a result, $P_{+}$ has two positive roots in terms of $y^2$, and hence four real roots in terms of $y$, two of them positive: 
\begin{equation}\label{rootspMstaro}
y^{\pm}_{P_+} =  \left[\frac{ \rho_{\rm max} }{12} \left( 7 + 2  \sqrt{ \frac{3}{ \pi} }D_S+ \frac{D^2_S}{ 2\pi}  \pm  \sqrt{ \Delta_{P_{+}}} \right) \right]^{1/2}, 
\end{equation}
and the other two negative, given by $- y^{\pm}_{P_+} $. Since $P_{+}$ tends to minus infinity for large absolute values of $y$, it immediately follows from the bound $\breve{s}^{({\rm s})}_{\rm B}/(8\pi a^2_{\rm B})\leq P_{+}(y)$ that, in the considered zeroth-order approximation in $A_S$ in which $y^2\approx V_S^{\rm B}$, the mass $\breve{s}^{({\rm s})}_{\rm B}$ is negative for values of the Starobinsky potential smaller than $(y_{P_+}^{-})^2$. It is possible to show that $(y_{P_+}^{-})^2 \in (0,\rho_{\rm max}/6]$, varying with the value of the parameter $D_S$. For instance, for $D_s=\sqrt{16 \pi/3}$, which is the usually employed value, we get $(y_{P_+}^{-})^2\approx  \rho_{\rm max} /18$, whereas in the case of a parameter equal to one in Planck units, i.e. $D_S=1$, we obtain $(y_{P_+}^{-})^2 \approx \rho_{\rm max} /9$. Hence, the mass at the bounce in the dressed metric formalism is negative in a region that in general extends beyond the sector of kinetic dominance in the inflaton energy density at the bounce, sector for which the value of the potential is negligibly small.

Finally, adopting a truncation of Eq. \eqref{dressedmassstaro} at first subdominant order in $A_S$ similar to the one that we carried out in the previous section for Eq. \eqref{hybridmassstaro}, we can improve our approximation for the estimation of an interval in which the mass is negative. At this linear order, we replace $y_{P_+}^{-}$ with $y_{P_+}^{-}+A_S y_{P_+}^{-(1)}$, where
\begin{equation}\label{y1dressstaro}
y_{P_+}^{-(1)} = \frac{ D_S^2 \rho_{\rm max}  + 8 \pi \left(7 \rho_{\rm max} - 12 (y_{P_+}^{-})^2 \right) +8 D_S \sqrt{3 \pi \rho_{\rm max}} \left(\rho_{\rm max} - 2 (y_{P_+}^{-})^2 \right) \left( \rho_{\rm max} - (y_{P_+}^{-})^2\right)^{-1/2}}{4  \left( D_S^2 \rho_{\rm max} + 14 \pi \rho_{\rm max} + 4 D_S \rho_{\rm max} \sqrt{3 \pi} + 24 \pi (y_{P_+}^{-})^2 \right)}.
\end{equation}

\section{Exponential potentials} \label{sec:expo}

Let us now analyze the scalar mass at the bounce for exponential potentials. The most relevant cases are the hyperbolic cosine potential $V_C$ and the purely exponential potential $V_e$, on which we will focus our discussion. At least in the context of geometrodynamics, all other linear combinations of an exponential and its inverse can be recast in the form of these two cases by means of a suitable redefinition of fields, as we commented in Sec. \ref{sec:mass}. On the other hand, as we anticipated in that section, we will only discuss the most appealing physical scenario of a positive potential (for any real value of the inflaton). In addition, we will allow that the only minimum of the hyperbolic cosine potential can be made equal to zero by including the possibility of an additive constant, therefore analyzing the new potential $\tilde{V}_C=V_C+E_C$. For convenience, we also contemplate the introduction of an additive constant $E_e$ in the purely exponential potential, although we will eventually make it vanish. In this way, we can use the similar notation $\tilde{V}_e=V_e+E_e$. Then, ignoring the label $C$ or $e$ in the potentials and in the parameters that appear in their definitions, we can use the compact notation $\tilde{V}=A \,F(D\phi)+E$, where the function $F$ is the hyperbolic cosine or the pure exponential, depending on the specific potential under analysis. Note that, according to our previous comments in this paragraph and in Sec. \ref{sec:mass}, the parameters $A$ and $D$ can be taken as positive, because of the parity behavior and positivity of the function $F$ in the two studied cases. Besides, in both cases we have
\begin{equation}
\tilde{V}_{,\phi\phi}=D^2 \tilde{V}- D^2 E.
\end{equation}
With these considerations, most of the analysis of the positivity or negativity of the mass at the bounce can be carried out simultaneously. 

\subsection{Exponential potentials in the hybrid formalism}

With the above notation and results, we can write the scalar mass at the bounce in the form
\begin{equation} \label{hybmassexpo}
\frac{s_{\rm B}^{({\rm s})}}{8\pi a_{\rm B}^2}= \frac{\rho_{\rm max}}{3} - \frac{D^{2}E}{8 \pi} + \left( 5  + \frac{D^{2}}{8 \pi} \right)  \tilde{V}^{\rm B}  - \frac{6} { \rho_{\rm max}} \left(\tilde{V}^{\rm B}\right)^2:= P_e(\tilde{V}^{\rm B}),
\end{equation} 
where we have called $P_e$ the resulting quadratic polynomial of the potential. Since the coefficient of the quadratic term is strictly negative, the polynomial is positive only in the interval delimited by its two roots, if they exist. Therefore, the mass turns out to be positive in the intersection of this interval with the allowed range of values of the potential at the bounce, which we recall that is always bounded from above by the maximum energy density $\rho_{\rm max}$ (because the kinetic contribution is positive). 

To analyze the properties of the roots of the polynomial $P_e(\tilde{V}^{\rm B})$, let us study its discriminant,
\begin{equation}\label{Deltahybmassexpo}
\Delta_{P_e}= \left( 5  + \frac{D^{2}}{8 \pi} \right) ^2 + 8 - \frac{ 3 D^{2} E}{\pi \rho_{\rm max}}.
\end{equation}
If this discriminant is strictly negative, the polynomial $P_e$ does not admit any real root, and then the scalar mass is negative at the bounce. A direct calculation shows that the opposite situation, with a positive discriminant, is found if and only if
\begin{equation}\label{positivedis}
E \leq \frac{  2112 \pi^2 + 80 \pi D^{2} + D^{4} }{192 \pi D^{2} }          \rho_{\rm max}.
\end{equation}
We notice that this inequality holds in fact for a vanishing constant $E$. In particular, this is the most relevant case for a purely exponential potential. For the hyperbolic cosine, if instead of the possibility $E=0$ we want that the minimum of the potential is set equal to zero, we must take $E=-A_C$ (we re-establish here the label $C$ to make clear that we refer to the case of the hyperbolic cosine). Then, recalling that $A_C$ is positive, we see that the above inequality is immediately satisfied. 

Let us admit that our condition on $E$ is verified, as in the cases of interest discussed above. As a consequence, the scalar mass at the bounce turns out to be strictly positive for potentials $\tilde{V}^{\rm B}\leq \rho_{\rm max}$ that belong to the interval delimited by the two roots $v^{\pm}_{P_e}$ of $P_e$. These roots are
\begin{equation}\label{rootshybexpo}
v^{\pm}_{P_e}= \frac{40 \pi + D^{2} \pm  \sqrt{ 2112 \pi^2  + (80-192 \tilde{E} ) \pi D^{2}  + D^{4} }}{96 \pi  }          \rho_{\rm max},
\end{equation}
where $\tilde{E}=E/ \rho_{\rm max}$. It is straightforward to check that the smallest root, $v^{-}_{P_e}$, is negative if and only if $\tilde{E}\leq 8\pi /(3 D^{2} )$. As required for consistency, this condition is stronger than inequality \eqref{positivedis}, found above for the existence of $v^{-}_{P_e}$. 
Clearly, this restriction is satisfied for vanishing $E$, as well as for $E=-A_C< 0$ in the case of the hyperbolic cosine potential\footnote{In a similar way, it is possible to show that, if we allowed the amplitude of the hyperbolic cosine contribution to be negative and we wanted that the interval with positive mass includes the zero value of the potential, we would have to restrict the study to parameters such that $A_C=-E_C > - 8\pi \rho_{\rm max}/(3 D^{2} )$.}. 

In any of these situations, we have proven that the scalar mass at the bounce in the hybrid formalism is positive in the interval of potentials
\begin{equation}
[0, {\rm Min}(v^{+}_{P_e},\rho_{\rm max})].
\end{equation}
Here, ${\rm Min}(\cdot,\cdot)$ stands for the minimum of its two arguments. Moreover, it is easy to see from Eq. \eqref{rootshybexpo} that we always have $v^{+}_{P_e}> 5 \rho_{\rm max} /12$ independently of the value of $E$. Obviously, this lower bound can be improved if this value is fixed. For instance, for vanishing $E$ one can check that our formulas reproduce those obtained for the Starobinsky potential in Subsec. \ref{sec:starobinsky} at dominant order in the amplitude of that potential, under the substitution of $D$ by $2D_S$.  Therefore, it is easy to realize that, when $E=0$, the lower bound on $v^{+}_{P_e}$ can be improved at least to $5 \rho_{\rm max} /6$. In any case, the interval of values of the exponential potentials in which the positivity of the mass is guaranteed clearly contains the sector of kinetic dominance at the bounce, sector in which the potential is negligibly small.

\subsection{Exponentials potential in the dressed metric formalism} \label{sec:cosh}

In the case of the dressed metric formalism, we follow the same procedure as with the Starobinksy potential, finding polynomic bounds for the mass at the bounce. This mass is now 
\begin{equation}
\frac{\breve{s}_{\rm B}^{(s)}}{8 \pi a_{\rm B}^{2}}=-\rho_{\max }-\frac{D^{2}E }{8 \pi }+\left(7 +\frac{D^{2}}{8 \pi }\right) \tilde{V}^{\rm B}-\frac{6}{\rho_{\text {max }}} \left(\tilde{V}^{\rm B}\right)^{2} \pm \sqrt{\frac{3}{2\pi \rho_{\rm max}}}   \frac{|\phi^{\prime}_{\rm B}|\tilde{V}_{,\phi}^{\rm B}}{a_{\rm B}},
\end{equation}
where we have contemplated the two possible values of the sign $\sigma_a$ in Eq. \eqref{Vformula}. Recall that the considered (positive) exponential potentials take values in the interval $0 \leq \tilde{V}^{\rm B} \leq\rho_{\rm max} $. In addition, the derivative of both the purely exponential potential and the hyperbolic cosine potential satisfies 
\begin{equation}
| \tilde{V}_{,\phi}^{\rm B} | \leq D ( \tilde{V}^{\rm B} - E ) =A D \, F( D \phi_{\rm B} ).
\end{equation} 
Note that $AD  F(D \phi_{\rm B}) \geq 0$ because $A$, $D$, and the function $F$ are positive. Employing these facts and the expression of the inflaton energy density \eqref{density} at the bounce, we can bound the term proportional to $\phi^{\prime}_{\rm B}$ in the mass by
\begin{equation}
\sqrt{\frac{3}{2\pi \rho_{\rm max}}}  \sqrt{2(\rho_{\text {max }}-\tilde{V}^{\rm B})} |\tilde{V}_{,\phi}^{\rm B}| \leq \sqrt{\frac{3}{\pi}}D(\tilde{V}^{\rm B} - E).
\end{equation}
Hence, we obtain the following upper bound on the mass: 
\begin{equation}
\label{def_P+}
\frac{\breve{s}_{\rm B}^{(s)}}{8 \pi G a_{\rm B}^{2}} \leq \left(-\rho_{\text {max }} -\sqrt{\frac{3}{\pi}}D E-\frac{D^{2} E}{8 \pi }\right) +\left(7+\sqrt{\frac{3}{\pi}}D+\frac{D^{2}}{8 \pi }\right)\tilde{V}^{\rm B}-\frac{6}{\rho_{\max }} \left(\tilde{V}^{\rm B}\right)^{2} :=\tilde{P}_+ (\tilde{V}^{\rm B}),
\end{equation}
where we have defined the polynomial $\tilde{P}_+(\tilde{V}^{\rm B})$ of the potential at the bounce.

Let us analyze the discriminant $\Delta_{\tilde{P}_+}$ of the polynomial $\tilde{P}_+$. A simple calculation shows that this discriminant is positive if and only if the following inequality holds:
\begin{equation}
E \leq  \frac{1600 \pi^{2}+896 \sqrt{3} \pi^{3 / 2} D +304  \pi D^{2} +16 \sqrt{3 \pi} D^{3} +D^{4}}{192 \pi(8 \sqrt{3 \pi}D+D^2)} \rho_{\max}:=M(D). 
\end{equation}
The function $M(D)$ is strictly positive for all positive values of $D$. Therefore, the condition on $E$ is satisfied if this constant is zero or strictly negative. This is precisely what happens in the two cases in which we are mainly interested, namely, a vanishing additive constant $E$ for any of the exponential potentials or a constant $E=-A_C< 0$ for the hyperbolic cosine potential.  

The polynomial $\tilde{P}_+$ has then two real roots, given by
\begin{equation}
r_{\pm}=\frac{\rho_{\max }}{12}\left(7+\sqrt{\frac{3}{\pi}}D +\frac{D^{2}}{8 \pi}\pm \sqrt{\Delta_{\tilde{P}_{+}}}\right).
\end{equation}
In the case $E= 0$, one immediately sees that 
\begin{equation}\label{discrdressexpo}
\left(7+\sqrt{\frac{3}{\pi}}D +\frac{D^{2}}{8 \pi}\right)^2 \geq \Delta_{\tilde{P}_{+}}.
\end{equation}
In fact, this follows straightforwardly from our discussion in Subsec. \ref{subsec:dressstarobinsky}, once one realizes that the expression of $\Delta_{\tilde{P}_{+}}$ for $E=0$ coincides with that of $\Delta_{P_{+}}$ in Eq. \eqref{DeltapMstaro} after replacing $D$ with $2D_S$. In the other interesting case for the hyperbolic cosine potential, namely $E=-A_C< 0$, a trivial calculation shows that the above inequality applies as well if the amplitude of the coshine contribution is not too large, namely
\begin{equation}\label{coshamplitude}
A_C\leq \frac{8\pi }{8\sqrt{3\pi}D+D^2}\, \rho_{\max}.
\end{equation}
In any of these situations, or more generally when Eq. \eqref{discrdressexpo} holds, the smallest root $r_{-}$ is positive. Since, at the bounce, $\tilde{P}_{+}$ bounds the mass from above and it is negative for all potentials equal to or lower than $r_{-}$, we can ensure the negativity of the scalar mass when $\tilde{V}^{\rm B} \in [0,r_{-}]$. This interval of values for the potential includes the sector in which the inflaton energy density is dominated by the kinetic contribution. To this extent, we can say that the results of Ref. \cite{mass} about the negativity of the scalar mass in the dressed metric formalism are valid as well for the studied exponential potentials. For the sake of an example, if $A_C$ is of the order of the phenomenologically preferred value for the parameter $A_S^2$ of the Starobinsky potential, let's say $A_C=10^{-12}\rho_{\rm max}$, we get $r_{-}\approx 0.139 \rho_{\rm max}$ when $E=-A_C$ and $D$ is equal to the Planck unit.

On the other hand, in consonance with our comments above, when the additive constant $E$ in the potential vanishes, the value of $r_{-}$ coincides with the squared root $(y_{P_+}^{-})^2$ of the Starobinsky potential with the replacement of $D$ with $2D_S$ [see Eqs. \eqref{DeltapMstaro} and \eqref{rootspMstaro}]. We can then apply the same analysis that we carried out in Subsec. \ref{subsec:dressstarobinsky} to estimate the value of $r_{-}$. Based on that analysis, in the absence of an additive constant in the potential we can ensure again the negativity of the scalar mass in a region which covers and generally extends beyond the sector of kinetic dominance where the potential is negligible at the bounce. 

\section{Conclusions} \label{sec:conclusion}

The dynamical equations that rule the propagation of cosmological perturbations in the preinflationary and inflationary epochs are modified by quantum geometry effects in LQC. We have considered two different approaches to this quantum description: the hybrid and the dressed metric formalisms. In absence of backreaction and for quantum states of the background that are peaked on effective trajectories, the main difference between the corresponding equations is found in the background-dependent mass that appears in them. This mass is different for scalar and tensor perturbations, but it is independent of the considered Fourier mode. Since the region with relevant LQC effects is very narrow around the bounce that replaces the cosmological singularity in the effective trajectories, we have studied the mass of the perturbations at this moment of the evolution. With this objective, we have generalized the analysis that was carried out in Ref. \cite{mass} in the case of a quadratic potential for the inflaton (or potentials with similar behaviors, including their first two derivatives). More specifically, we have considered two other types of potentials. The first one is the Starobinsky potential, which has received considerable attention in cosmology in the last years because it provides a remarkably good fit of the observations of the CMB within the framework of slow-roll inflation in general relativity. The second type are exponential potentials. These potentials have a more academic interest and, at least in geometrodynamics, lead to solvable models in the classical and quantum theories. The two prototypical examples on which we have focused our analysis are the purely exponential potential and the hyperbolic cosine potential, both of them defined as positive functions of the inflaton. In the case of the coshine, we have also allowed the addition of a constant to the potential in order to make its minimum equal to zero.

The positivity or negativity of the mass of the perturbations is of utmost importance, especially at the bounce if one finds that it is natural to fix there the initial conditions for the cosmological inhomogeneities and anisotropies. If the mass is negative, there exist Fourier modes of small wavenorm for which the dynamical equations are in fact of elliptic signature, rather than of hyperbolic nature. This implies that the typical mode solutions become exponentials instead of oscillatory functions. This indicates the possible appearance of instabilities. It also affects any attempt to find a natural set of positive frequency solutions, which one would associate to a physically motivated choice of vacuum state. Moreover, it is an obstruction for the implementation of certain constructions of this vacuum, as it is the case of the adiabatic states \cite{adiabatic,adiabaticLR}, since there would exist modes for which the adiabatic approximation\footnote{The positivity of the mass does not even guarantee by itself the validity of the WKB approximation, at the lowest nontrivial adiabatic order. See, e.g., Ref. \cite{Ref1}.} would break down \cite{Diag}. It is worth emphasizing that, if the vacuum state must be optimally adapted to the background evolution and capture the quantum geometry effects, it is most reasonable to select it by criteria based on first principles that impose conditions at the bounce, or in a narrow region around it\footnote{Since the evolution of the perturbations is a one-to-one map, one can equivalently impose conditions to select a vacuum at any other time. For instance, it has been suggested that they can be imposed on the asymptotic past. The physical principles supporting the choice of those conditions should admit a neat relation with the presence of quantum geometry effects. On the other hand, our analysis of the behavior of the mass of the perturbations can be generalized to times other than the bounce. For a quadratic potential, a discussion of this mass in the asymptotic past was carried out in Ref. \cite{AGG}, adopting an alternative regularization of the Hamiltonian constraint in homogeneous LQC.}. According to Ref. \cite{analyticvacuum}, criteria of this kind are expected to avoid large oscillations in the primordial power spectra and lead to suppression at scales that, when evolved backwards to the bounce, are of Planck order\footnote{In fact, suppresion at these scales is found, e.g., for the vacuum put forward by Ashtekar and Gupt in the dressed metric formalism \cite{AsG1,AsG2}, although with superimposed oscillations in the spectrum that may produce some extra power on average. On the other hand, other proposals, like that of Ref. \cite{Blas}, do not display the desired suppression scale.}. Actually, it has been argued \cite{Neves} that states with the properties of the vacuum studied in Ref. \cite{analyticvacuum} are of Hadamard type, a claim that brings into focus the question of whether this vacuum may be reached in a certain limit from high-order adiabatic states. Owing to these reasons, the positivity of the mass is not only a convenient property, but rather a fundamental ingredient for the application and extension of many of the techniques that have been developed in quantum field theory in curved spacetimes.  

For the mass of the tensor perturbations, the discussion presented in Ref. \cite{mass} for the quadratic potential can be extended immediately to the case of the Starobinsky and the exponential potentials, confirming the positivity of the mass at the bounce for the hybrid formalism and its negativity for the dressed metric formalism, in both cases for an ample interval of values of the potential that contains the region corresponding to kinetic dominance. In fact, these results apply as well to generic potentials, beyond those studied in this work, and therefore are extremely solid. We have then analyzed in detail the mass for the scalar perturbations. In the hybrid formalism, we have shown that the positivity of this mass is still true for all the studied potentials: the Starobinsky, the purely exponential, and the hyperbolic cosine potentials. For the Starobinsky potential, and at dominant order in the phenomenologically small value of its amplitude, we have proven the positivity for values of the potential at the bounce equal or smaller than $5\rho_{\max}/6$. Similarly, for an exponential potential, its value at the bounce can be at least as large as $5\rho_{\max}/12$, including the case of a coshine with an added constant that sets the global minimum to zero. These results demonstrate the robustness of the positivity of the mass, for different behaviors of the inflaton potential and for values that are not negligible compared to the kinetic contribution, although the range of these values depends on the specific potential.
 
Finally, we have confirmed the negativity of the mass of the scalar perturbations in the dressed metric formalism, both for the Starobinsky and the exponential potentials. In this last class of potentials, we have also discussed a hyperbolic cosine with an added constant that shifts the minimum to a vanishing value. In all these cases, there exists an interval of values of the potential at the bounce for which the scalar mass is negative, and such that the interval contains the region of kinetically dominated energy density, around the zero value of the potential. The exact interval in which this happens, nonetheless, depends on the parameters of the model, and it is generally not possible to find a common interval that is valid for all values of those parameters in the dressed metric approach.
\acknowledgments
	
The authors are grateful to A. Garc\'{\i}a-Quismondo and B. Elizaga Navascu\'es for fruitful discussions. This work has been supported by Project. No. MICINN PID2020-118159GB-C41.

\end{document}